\documentclass{elsart}

\usepackage{epsfig}
\usepackage{cite}
\usepackage{amssymb}

\begin{document}

\begin{frontmatter}


  
  \begin{flushright}
    Cavendish-HEP-2004/11\\
    DAMTP-2004-26\\
    DESY 04-085
  \end{flushright}

  \title{The Gluon Green's Function in N~=~4 Supersymmetric Yang--Mills Theory}


\author[CAVENDISH,DAMTP]{Jeppe R.~Andersen}
\ead{andersen@hep.phy.cam.ac.uk},
\author[HAMBURG]{Agust{\'\i}n Sabio Vera}\footnote{Alexander von Humboldt Postdoctoral Research Fellow}
\ead{sabio@hep.phy.cam.ac.uk}

\address[CAVENDISH]{Cavendish Laboratory, University of Cambridge, Madingley Road, CB3 0HE, Cambridge, UK}
\address[DAMTP]{DAMTP, Centre for Mathematical Sciences, Wilberforce Road, CB3 0WA, Cambridge, UK}
\address[HAMBURG]{II. Institut f{\"u}r Theoretische Physik, Universit{\"a}t Hamburg, Luruper Chaussee 149, 22761~Hamburg, Germany}

\begin{abstract}
  The high energy limit of scattering amplitudes in the N~=~4 supersymmetric
  Yang--Mills theory is studied by solving the corresponding BFKL equation in
  the next--to--leading approximation.  The gluon Green's function is
  analysed using a newly proposed method
suitable for investigating the contribution from higher conformal spins.  
From this new approach complete agreement is obtained with the
  results of Kotikov and Lipatov on conformal spins and angular dependence.
\end{abstract}

%\begin{keyword}

%\PACS 
%\end{keyword}
\end{frontmatter}

% main text

\section{Introduction}
In recent years there has been a large interest in the study of the
next--to--leading (NLL) corrections to the Balitsky--Fadin--Kuraev--Lipatov~\cite{FKL} 
(BFKL) equation in QCD~\cite{running,Fadin:1998py,Ciafaloni:1998gs,NLLpapers}. 
The calculation of the NLL corrections was extended
to supersymmetric gauge theories in
Ref.~\cite{Kotikov}. In that work it was
shown how in a N~=~4 SYM theoretical playground the kernel of the BFKL
equation is simplified. In particular, the analyticity of the eigenvalues in
terms of the conformal spins allowed the study of the connection between the
DGLAP and BFKL equations in this model, this being possible 
mainly due to the lack of coupling constant renormalisation. 
In Ref.~\cite{Kotikov:2003fb} the
anomalous dimension matrix of the Wilson twist--2 operators in this maximally 
supersymmetric theory in four dimensions 
was calculated at two loops. Very recently these results were extended to
three loops in Ref.~\cite{Kotikov:2004er} using the calculation for the
non--singlet case first obtained in Ref.~\cite{Moch:2004pa}, and in agreement, 
in the supersymmetric limit, 
with the results for the singlet case derived in Ref.~\cite{Vogt:2004mw}.

In the present study, the method for solving the BFKL equation at NLL accuracy 
developed in Ref.~\cite{Andersen:2003an,Andersen:2003wy} is applied to N~=~4
SYM. This will serve two purposes. Firstly, since the method solves the BFKL
equation with full angular information, it can be used to test the results in
the literature \cite{Kotikov} for the dependence of the
eigenvalues of the kernel on conformal spins. Secondly, due to the
conformal invariance of the theory also at higher orders, the analytic solution
to the NLL BFKL equation is known for N~=~4 SYM, and therefore it is possible
to directly test the results obtained using the completely independent method
described in this work. The results in the literature for the solution of the
N~=~4 SYM BFKL equation at NLL in terms of conformal eigenfunctions are
reviewed in the next section. In Sec.~\ref{TranSol} the NLL BFKL equation is
solved directly in momentum space using the iterative method of
Ref.\cite{Andersen:2003an,Andersen:2003wy}.  Sec.~\ref{sec:analys-traj-emiss}
is devoted to a study of the numerical structure of the trajectory and the
real emission kernel arising in this approach. In
Sec.~\ref{sec:study-gluon-greens} the results obtained in the two approaches
are compared and the conclusions are presented.

\section{The BFKL equation and its solution in Mellin space}
\label{MellSol}

In the BFKL formalism the high energy limit of a scattering process factorizes as 
\begin{eqnarray}
\sigma(s) &=&\int 
\frac{d^2 \vec{k}_a}{\vec{k}_a^2}
\int \frac{d^2 \vec{k}_b}{\vec{k}_b^2} ~\Phi(\vec{k}_a) ~\Phi'(\vec{k}_b)
~f \left(\vec{k}_a,\vec{k}_b, {\rm Y} \equiv \ln{\frac{s}{s_0}}\right),
\end{eqnarray}
where $s_0 = |\vec{k}_a| |\vec{k}_b|$ is the Regge scale.  The energy
dependence is determined by the universal
process--independent gluon Green's function $f$.  The impact factors, 
$\Phi, \Phi'$, depend on the process under study.  In the Regge--limit
  of large centre of mass energy and fixed momentum transfer the
effective degrees of freedom are the transverse momenta $\vec{k}_{a,b}$ of
the exchanged gluons.  The dynamics of these processes is then
two--dimensional evolving with a time variable Y. The evolution of the gluon
Green's function with Y is governed by the BFKL equation. This equation is
traditionally written in terms of a Mellin transform of Y, i.e.
\begin{eqnarray}
f \left(\vec{k}_a,\vec{k}_b, {\rm Y}\right) 
&=& \frac{1}{2 \pi i}
\int_{a-i \infty}^{a+i \infty} d\omega ~ e^{\omega {\rm Y}} f_{\omega} 
\left(\vec{k}_a ,\vec{k}_b\right).
\label{Mellin}
\end{eqnarray}
With such transformation the equation reads
\begin{eqnarray}
\omega f_\omega \left(\vec{k}_a,\vec{k}_b\right) &=& \delta^{(2)} 
\left(\vec{k}_a-\vec{k}_b\right) + \int d^{2}\vec{k}' ~
\mathcal{K}\left(\vec{k}_a,\vec{k}'\right)f_\omega \left(\vec{k}',\vec{k}_b 
\right).
\label{first}
\end{eqnarray}
The inhomogeneous term of this integral equation corresponds 
to a single gluon
exchange and the kernel (in dimensional regularization, $D = 4 + 2 \epsilon$)
\begin{eqnarray}
\mathcal{K}\left(\vec{k}_a,\vec{k}\right) = 2 \,\omega^{(\epsilon)}\left(\vec{k}_a^2\right) \,\delta^{(2+2\epsilon)}\left(\vec{k}_a-\vec{k}\right) + \mathcal{K}_r\left(\vec{k}_a,\vec{k}\right)
\end{eqnarray}
contains the gluon Regge trajectory~\cite{Kotikov}, 
\begin{eqnarray}
2 \, \omega^{(\epsilon)} \left(\vec{q}^2\right) &=& - a \left( \frac{1}{\epsilon} + \ln{\frac{\vec{q}^2}{\mu^2}}\right) \nonumber \\
&&- \frac{a^2}{8} \left[\left(\frac{1}{3}-2 \zeta(2)\right)\left(\frac{1}{\epsilon}+2 \ln{\frac{\vec{q}^2}{\mu^2}}\right)-\frac{8}{9}+2 \zeta(3)\right],
\end{eqnarray}
with the coupling $a = \frac{{\overline g}^2 N_c}{4 \pi^2}$, which is not
running in N~=~4 SYM. The kernel also includes two contributions from the real
emissions:
\begin{eqnarray}
\mathcal{K}_r &=& {\mathcal K}_r^{(\epsilon)} + \widetilde{\mathcal K}_r
\end{eqnarray}
where~\cite{Kotikov}
\begin{eqnarray}
\mathcal{K}_r^{(\epsilon)} \left(\vec{q},\vec{q}+\vec{k}\right) &=&
\frac{a \, \mu^{-2 \epsilon}}{\pi^{1+\epsilon} \Gamma(1-\epsilon)} \frac{1}{\vec{k}^2} \nonumber\\
&\times&\left\{1+\frac{a}{4} \left(\frac{\vec{k}^2}{\mu^2}\right)^\epsilon \left[\frac{1}{3}- 2 \, \zeta(2)+\epsilon \left(-\frac{8}{9}+14 \, \zeta(3)\right)\right]\right\} 
\end{eqnarray}
and~\cite{Kotikov} 
\begin{eqnarray}
\widetilde{\mathcal{K}}_r \left(\vec{q}, \vec{p}\right) &=& 
\frac{a^2}{4 \pi} 
\left\{-\frac{1}{(\vec{q}-\vec{p})^2}
\ln^2{\frac{\vec{q}^2}{\vec{p}^{2}}} \right. \nonumber\\
&&\hspace{-3cm}
+\frac{2(\vec{q}^2-{\vec{p}}^2)}{(\vec{q}-{\vec{p}})^2(\vec{q}+{\vec{p}})^2} 
\left(\frac{1}{2}\ln{\frac{\vec{q}^2}{{\vec{p}}^2}}
\ln{\frac{\vec{q}^2 {\vec{p}}^2 (\vec{q}-{\vec{p}})^4}
{(\vec{q}^2+{\vec{p}}^2)^4}}
+ \left( \int_0^{- \frac{\vec{q}^2}{{\vec{p}}^2}} -
\int_0^{- \frac{{\vec{p}}^2}{\vec{q}^2}} \right) 
dt \frac{\ln(1-t)}{t}\right)\nonumber\\
&&\hspace{-3cm}\left.-\left(1-\frac{(\vec{q}^2-{\vec{p}}^2)^2}{(\vec{q}-{\vec{p}})^2 (\vec{q}+{\vec{p}})^2}\right) 
\left( \left( \int_0^1 
-\int_1^\infty \right) dz \frac{1}{({\vec{p}}-z \vec{q})^2}
\ln{\frac{(z \vec{q})^2}{{\vec{p}}^2}}\right)\right\}.
\label{KTILDE}
\end{eqnarray}

In N~=~4 SYM the BFKL kernel respects conformal symmetry even at NLL
accuracy, and the eigenfunctions do not change compared to the leading logarithmic (LL) 
ones. The solution to this equation can therefore be found using the expansion
on the known eigenfunctions
\begin{eqnarray}
f\left(\vec{k}_a,\vec{k}_b,Y\right) &=&
\frac{1}{\pi |\vec{k}_a| |\vec{k}_b|} 
\sum_{n=-\infty}^{\infty} \int \frac{d\omega }{2 \pi i}\, e^{\omega Y}
\int \frac{d \gamma}{2 \pi i} 
\left(\frac{\vec{k}_a^2}{\vec{k}_b^2}\right)^{\gamma-\frac{1}{2}}
\frac{e^{i n \theta}}{\omega - \omega_n (a,\gamma)},
\label{expansion}
\end{eqnarray}
with $\theta$ the angle defined by the $\vec{k}_a$ and $\vec{k}_b$ 
transverse momenta. The Fourier transform in angles is characterized by 
the so--called conformal spins $n$. Given that the coupling is fixed 
it is possible to fully diagonalize the 
BFKL kernel by simply acting on the LL eigenfunctions, i.e., in the 
$\overline{\rm MS}$ scheme we have~\cite{Kotikov}
\begin{eqnarray}
\int d^{2} \vec{q} \, \mathcal{K} \left(\vec{k},\vec{q}\right) 
\left(\frac{\vec{q}^2}{\vec{k}^2}\right)^{\gamma-1} 
e^{i n \theta} &=& \omega_n (a,\gamma) = \xi^{\rm MS} \chi(|n|,\gamma) + \eta + \Omega(|n|,\gamma).
\label{eigenvaluesf}
\end{eqnarray}
In this expression we have defined
\begin{eqnarray}
\xi^{\rm MS} &\equiv& a + a^2 \left(\frac{1}{12}-\frac{\zeta(2)}{2} \right)
\,,\,\,\,\eta ~\equiv~ a^2 \, \frac{3}{2} \, \zeta (3)
\label{xieta}
\end{eqnarray}
and 
\begin{eqnarray}
\chi(n,\gamma) &=& 2 \Psi(1) - \Psi\left(\gamma+\frac{n}{2}\right)-\Psi\left(1-\gamma+\frac{n}{2}\right), \\
&&\hspace{-2.7cm}\Omega(n,\gamma) ~=~ \frac{a^2}{4} \left[\Psi''\left(\gamma+\frac{n}{2}\right)+
\Psi''\left(1-\gamma+\frac{n}{2}\right)-2 \Phi(n,\gamma)-2 \Phi(n,1-\gamma)\right] \\
\Phi(n,\gamma) &=& \sum_{k=0}^{\infty} \frac{(-1)^{(k+1)}}{k+\gamma+\frac{n}{2}}
\left[\frac{}{}\Psi'(k+n+1)-\Psi'(k+1)\right.\nonumber\\
&&\hspace{2cm}+(-1)^{(k+1)} \left(\beta'(k+n+1)+\beta'(k+1)\right)\nonumber\\
&&\hspace{2cm}-\left.\frac{1}{k+\gamma+\frac{n}{2}}\left(\Psi(k+n+1)-\Psi(k+1)\right)\right],\\
\beta'(z) &=& \frac{1}{4}\left[\Psi'\left(\frac{1+z}{2}\right)-\Psi'\left(\frac{z}{2}\right)\right].
\end{eqnarray}
We have checked that switching to the gluon--bremsstrahlung (GB) scheme
\begin{eqnarray}
a_{\rm GB} &=& a 
+ \left(\frac{1}{12}-\frac{\zeta(2)}{2}\right) a^2, \, \, \, \, \,
\xi^{\rm GB} ~=~ a_{\rm GB},
\end{eqnarray}
or to the dimensional reduction (DRED) scheme, which respects SUSY~\cite{Kotikov},
\begin{eqnarray}
a_{\rm DRED} &=& a + \frac{1}{12} a^2, \, \, \, \, \, 
\xi^{\rm DRED} ~=~ a_{\rm DRED} - \frac{\zeta(2)}{2} {a_{\rm DRED}}^2,
\end{eqnarray}
the results of this work do not change qualitatively. 

The analyticity of these expressions, obtained by Kotikov and Lipatov
in~\cite{Kotikov}, for the dependence of the eigenvalues of the NLL BFKL
kernel on the conformal spins is very important. It allows the analytic
continuation to negative $|n|$ and to find the connection between the BFKL
equation and DGLAP in the N~=~4 supersymmetric gauge theory. One of the
objectives of the present work will be to confirm that these results for the
conformal spins are correct using a completely orthogonal method of solution
of the BFKL equation, which is developed in the next section.

\section{The solution directly in transverse--momentum space}
\label{TranSol}

The starting point of the new method of solution is Eq.~(\ref{first}) with 
the infrared divergences regularized in dimensional regularization. 
In the real emission kernel there are contributions 
which will lead to $\epsilon$ poles after phase space integration, 
$\mathcal{K}_r^{(\epsilon)}$, and others 
which will be finite, $\mathcal{K}_r$, i.e.
\begin{eqnarray}
\omega f_\omega \left(\vec{k}_a,\vec{k}_b\right) &=& \delta^{(2+2\epsilon)}
\left(\vec{k}_a-\vec{k}_b\right) \nonumber \\
&+& \int d^{2+2\epsilon}\vec{k} \, 2 \,
\omega^{(\epsilon)} \left(\vec{k}_a^2\right) \delta^{(2+2\epsilon)} 
\left(\vec{k}_a-\vec{k}\right) f_\omega \left(\vec{k},\vec{k}_b\right) \nonumber\\
&+& \int d^{2+2\epsilon}\vec{k} \, \mathcal{K}_r^{(\epsilon)} \left(\vec{k}_a,\vec{k}_a+\vec{k}\right) f_\omega \left(\vec{k}_a+\vec{k},\vec{k}_b\right) \nonumber\\
&+& 
\int d^{2+2\epsilon}\vec{k} \, \widetilde{\mathcal{K}}_r \left(\vec{k}_a,\vec{k}_a+\vec{k}\right) f_\omega \left(\vec{k}_a+\vec{k},\vec{k}_b\right).
\end{eqnarray}
In order to explicitly show the cancellation of the infrared $\epsilon$ 
divergencies we introduce a phase space slicing parameter, $\lambda$, in the 
integral over real emission, and make use of the approximation
\begin{eqnarray}
f_\omega \left(\vec{k}+\vec{k}_a,\vec{k}_b \right) 
&\simeq& f_\omega \left(\vec{k}+\vec{k}_a,\vec{k}_b \right) 
\theta\left(\vec{k}^2 - \lambda^2\right) +
f_\omega \left(\vec{k}_a,\vec{k}_b \right)
\theta\left(\lambda^2-\vec{k}^2\right).
\end{eqnarray}
This is valid when $\lambda$ is small compared to $k_a$.  Hence the
equation now reads
\begin{eqnarray}
\omega f_\omega \left(\vec{k}_a,\vec{k}_b\right) &=& \delta^{(2+2\epsilon)}
\left(\vec{k}_a-\vec{k}_b\right) \\
&&\hspace{-3cm}+ \left\{2 \, 
\omega^{(\epsilon)} \left(\vec{k}_a^2\right) + \int d^{2+2\epsilon}\vec{k} \, \mathcal{K}_r^{(\epsilon)} \left(\vec{k}_a,\vec{k}_a+\vec{k}\right) \theta\left(\lambda^2-\vec{k}^2\right) \right\}f_\omega \left(\vec{k}_a,\vec{k}_b\right)\nonumber\\
&&\hspace{-3cm}+ \int d^{2+2\epsilon}\vec{k} \, \left\{ \mathcal{K}_r^{(\epsilon)} \left(\vec{k}_a,\vec{k}_a+\vec{k}\right) \theta\left(\vec{k}^2-\lambda^2\right) + \widetilde{\mathcal{K}}_r \left(\vec{k}_a,\vec{k}_a+\vec{k}\right) \right\} f_\omega \left(\vec{k}_a+\vec{k},\vec{k}_b\right). \nonumber
\end{eqnarray}
We have performed the integration over phase space for those emissions below 
the infrared cut--off $\lambda$, obtaining the result
\begin{eqnarray}
\int d^{2+2\epsilon}\vec{k} \, \mathcal{K}_r^{(\epsilon)} \left(\vec{q},\vec{q}+\vec{k}\right) \theta\left(\lambda^2-\vec{k}^2\right) &=& \frac{a}{\Gamma(1+\epsilon)\Gamma(1-\epsilon)}
\frac{1}{\epsilon} \left(\frac{\lambda^2}{\mu^2}\right)^\epsilon \nonumber\\
&&\hspace{-4.4cm} \times \left\{1+\frac{a}{8} \left(\frac{\lambda^2}{\mu^2}\right)^\epsilon \left[\frac{1}{3}-2\,\zeta(2)+\epsilon \left(-\frac{8}{9}+14\,\zeta(3)\right)\right]\right\}. 
\end{eqnarray}
Now it is possible to show how the poles cancel by calculating the gluon 
Regge trajectory in our regularisation scheme, i.e.
\begin{eqnarray}
\omega_0 \left({\vec{q}}^2,\lambda\right) &\equiv&\lim_{\epsilon \to 0} \left\{ 2\, \omega^{(\epsilon)}\left(\vec{q}^2\right) + \int d^{2+2\epsilon}\vec{k} \,
\mathcal{K}_r^{(\epsilon)} \left(\vec{q},\vec{q}+\vec{k}\right) 
\theta \left(\lambda^2-\vec{k}^2\right) \right\} \nonumber\\
&=& - a \left\{ \ln{\frac{\vec{q}^2}{\lambda^2}}
+ \frac{a}{4} \left[\left(\frac{1}{3}- 2 \xi(2) \right)\ln{\frac{\vec{q}^2}{\lambda^2}}- 6 \, \zeta(3) \right]\right\} \nonumber\\
&\equiv& - \xi^{\rm MS} \ln{\frac{\vec{q}^2}{\lambda^2}} + \eta, 
\label{TRAJ}
\end{eqnarray}
where $\xi$ and $\eta$ coincide with those defined in Eq.~(\ref{xieta}). The 
corresponding real emission part will be of the form
\begin{eqnarray}
\lim_{\epsilon \to 0} \int d^{2+2\epsilon}\vec{k} \, \mathcal{K}_r^{(\epsilon)} \left(\vec{k}_a,\vec{k}_a+\vec{k}\right) \theta\left(\vec{k}^2-\lambda^2\right) f_\omega \left(\vec{k}_a+\vec{k},\vec{k}_b\right) &=& \nonumber\\
&&\hspace{-6cm} \int d^{2}\vec{k} \, \frac{1}{\pi \vec{k}^2} \, \xi^{\rm MS} \, \theta\left(\vec{k}^2-\lambda^2\right) f_\omega \left(\vec{k}_a+\vec{k},\vec{k}_b \right).
\end{eqnarray}
Finally the N=4 SYM NLL BFKL equation can be expressed as
\begin{eqnarray}
\left(\omega - \omega_0\left(\vec{k}_a^2,\lambda^2\right)\right) f_\omega \left(\vec{k}_a,\vec{k}_b\right) &=& \delta^{(2)} \left(\vec{k}_a-\vec{k}_b\right)
\nonumber\\
&&\hspace{-5cm}+ \int d^2 \vec{k} \left(\frac{1}{\pi \vec{k}^2} \, \xi^{\rm MS} \, \theta\left(\vec{k}^2-\lambda^2\right)+\widetilde{\mathcal{K}}_r \left(\vec{k}_a,\vec{k}_a+\vec{k}\right)\right)f_\omega \left(\vec{k}_a+\vec{k},\vec{k}_b\right). 
\label{SBFKLf}
\end{eqnarray}
Note that the treatment of the kernel is with its full angular dependence,
i.e. without angular averaging over the angle between $\vec{k}_a$ and
$\vec{k}_b$. It is therefore possible to extract the contribution to the solution from all 
conformal spins.

Eq.~(\ref{SBFKLf}) will be solved by iteration, generalising the
  procedure of Ref.\cite{Kwiecinski:1996fm,Schmidt:1996fg,Orr:1997im}. In
order to do so it is useful to introduce the notation
\begin{eqnarray}
\widehat{\mathcal{K}}_r \left(\vec{k}_a, \vec{k}_a+\vec{k},\lambda\right) 
&\equiv& \frac{1}{\pi \vec{k}^2} \xi^{\rm MS} \, \theta\left(\vec{k}^2-\lambda^2\right)
+\widetilde{\mathcal{K}}_r \left(\vec{k}_a,\vec{k}_a+\vec{k}\right). 
\end{eqnarray}
The $\omega$ dependence can go to the 
denominator of the right hand side of the equation and iterate, generating in 
this way multiple poles in the complex $\omega$ space, i.e.
\begin{eqnarray}
f_{\omega} \left(\vec{k}_a ,\vec{k}_b\right) &=& 
\frac{\delta^{(2)} \left(\vec{k}_a - \vec{k}_b \right)}{\omega - {\omega}_0 \left(\vec{k}_a^2,\lambda \right)}\\
&&\hspace{-1cm}+ \int d^2 \vec{k}_1 
\frac{\widehat{\mathcal{K}}_r
\left(\vec{k}_a,\vec{k}_a+\vec{k}_1,\lambda \right)}
{\omega - {\omega}_0 \left(\vec{k}_a^2,\lambda \right)} 
\frac{\delta^{(2)} \left(\vec{k}_a +\vec{k}_1 - \vec{k}_b\right)}
{\omega - {\omega}_0 \left(\left(\vec{k}_a+\vec{k}_1\right)^2,\lambda \right)} \nonumber\\
&&\hspace{-1cm}+ \int d^2 \vec{k}_1 
\frac{\widehat{\mathcal{K}}_r \left(\vec{k}_a,\vec{k}_a+\vec{k}_1,\lambda \right)}
{\omega - {\omega}_0 \left(\vec{k}_a^2,\lambda \right)}
 \int d^2 \vec{k}_2 
\frac{\widehat{\mathcal{K}}_r
\left(\vec{k}_a + \vec{k}_1,\vec{k}_a+\vec{k}_1+\vec{k}_2,\lambda \right)}
{\omega - {\omega}_0 \left(\left(\vec{k}_a+\vec{k}_1\right)^2,\lambda \right)} \nonumber\\
&&\hspace{1cm}\times
\frac{\delta^{(2)} \left(\vec{k}_a +\vec{k}_1+\vec{k}_2 - \vec{k}_b\right)}
{\omega - {\omega}_0 \left(\left(\vec{k}_a+\vec{k}_1+\vec{k}_2\right)^2,\lambda \right)} \nonumber\\
&&\hspace{-1cm}+ \cdots \nonumber
\end{eqnarray}
It is now possible to invert the Mellin transform as in Eq.~(\ref{Mellin})
and go back to energy space, with the final compact expression for the gluon
Green's function being
\begin{eqnarray}
  f(\vec{k}_a ,\vec{k}_b, {\rm Y})  &=&
  \exp{\left(\omega_0 \left(\vec{k}_a^2,\lambda\right) {\rm Y} \right)}
  \delta^{(2)} (\vec{k}_a - \vec{k}_b) \nonumber\\
  &&\hspace{-3cm}
+\sum_{n=1}^{\infty} \Bigg[\prod_{i=1}^{n} \int d^2 \vec{k}_i \int_0^{y_{i-1}}
  d y_i \left[\frac{\theta\left(\vec{k}_i^2 - \lambda^2\right)}{\pi \vec{k}_i^2} \, \xi^{\rm MS} \, +\widetilde{\mathcal{K}}_r \left(\vec{k}_a+\sum_{l=0}^{i-1}\vec{k}_l,
\vec{k}_a+\sum_{l=1}^{i}\vec{k}_l\right)\right]\nonumber\\
&&\hspace{-1cm}\times \exp\left(
\omega_0\left(\left(\vec{k}_a+\sum_{l=1}^{i-1}
  \vec{k}_l\right)^2,\lambda\right) (y_{i-1}-y_i)\right)\Bigg]\nonumber\\
&&\hspace{-1cm}\times\exp\left(
\omega_0\left(\left(\vec{k}_a+\sum_{l=1}^{n}
  \vec{k}_l\right)^2,\lambda\right) y_n\right) \delta^{(2)} \left(\sum_{l=1}^{n}\vec{k}_l 
+ \vec{k}_a - \vec{k}_b \right),
\end{eqnarray}
where the notation $y_0 \equiv Y$ has been used. We note that the solution has
been expressed as the phase space integral of a product of effective emission
vertices connected with no--emission probabilities.

Before proceeding further in the numerical study of the solution we present 
an analysis of the trajectory and real emission kernel in the next section.

\section{Analysis of the trajectory and emission kernel}
\label{sec:analys-traj-emiss}
\begin{figure} 
\centerline{\psfig{file=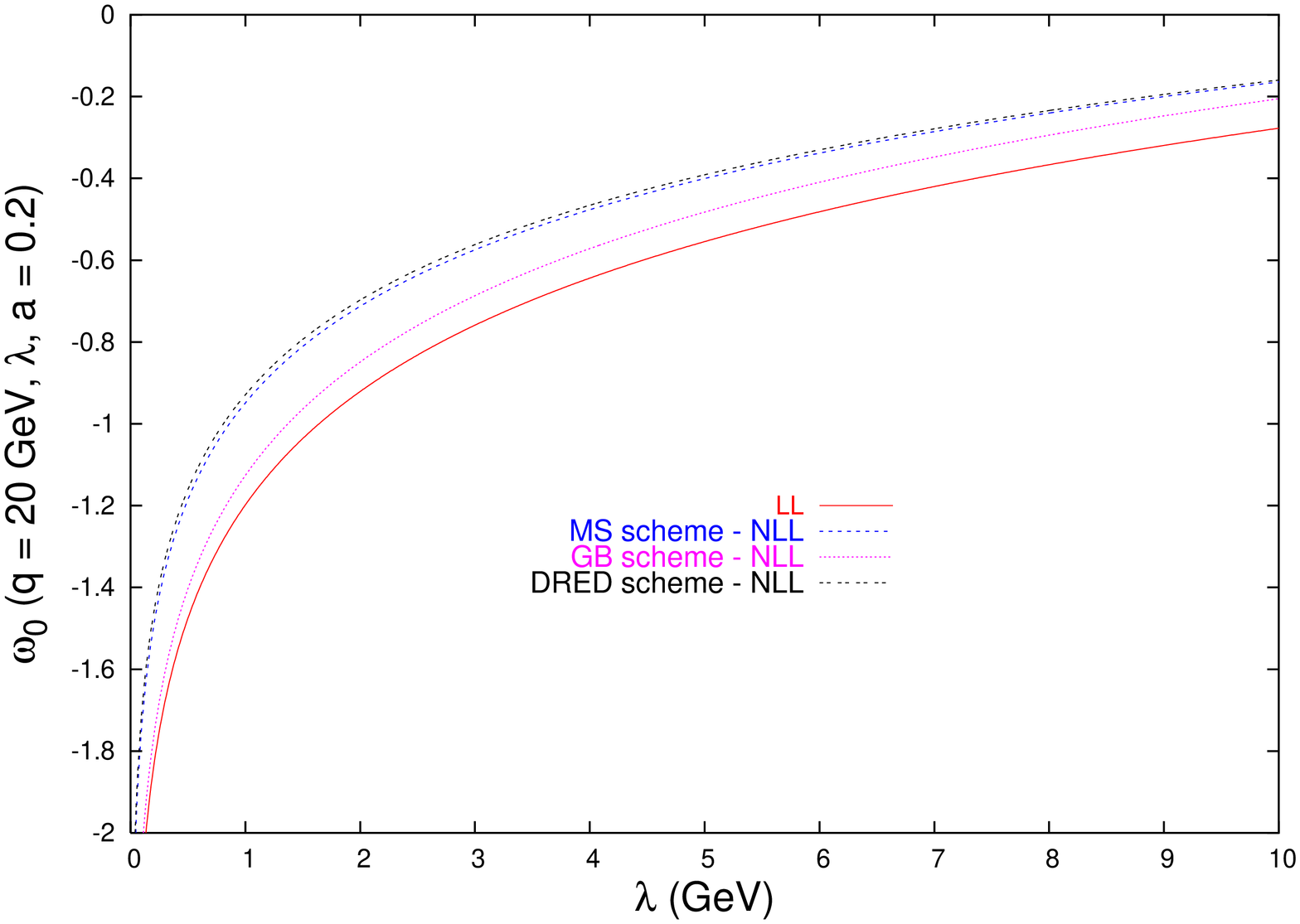,width=4.8cm,angle=-90}\psfig{file=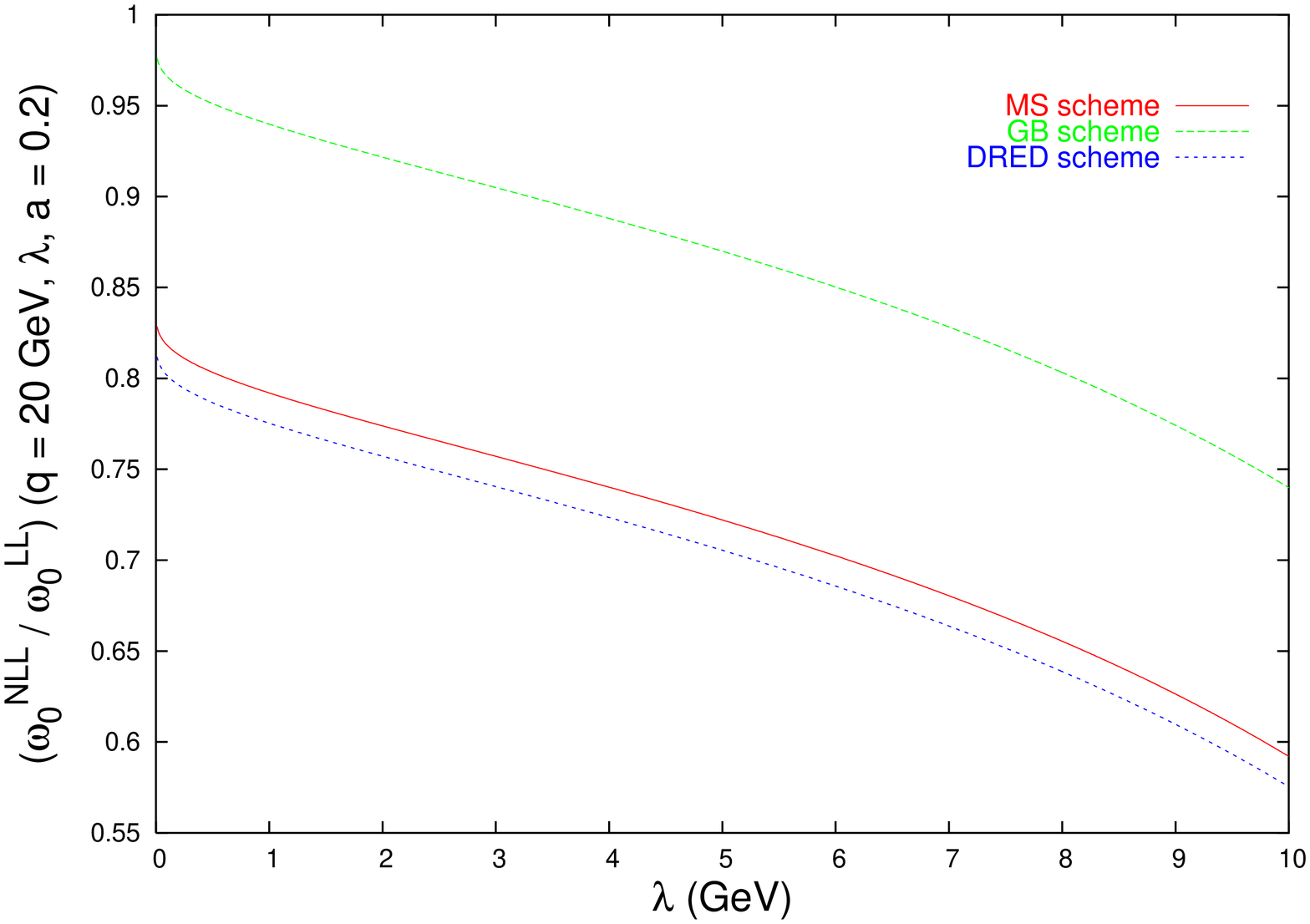,width=4.8cm,angle=-90}}
\centerline{\psfig{file=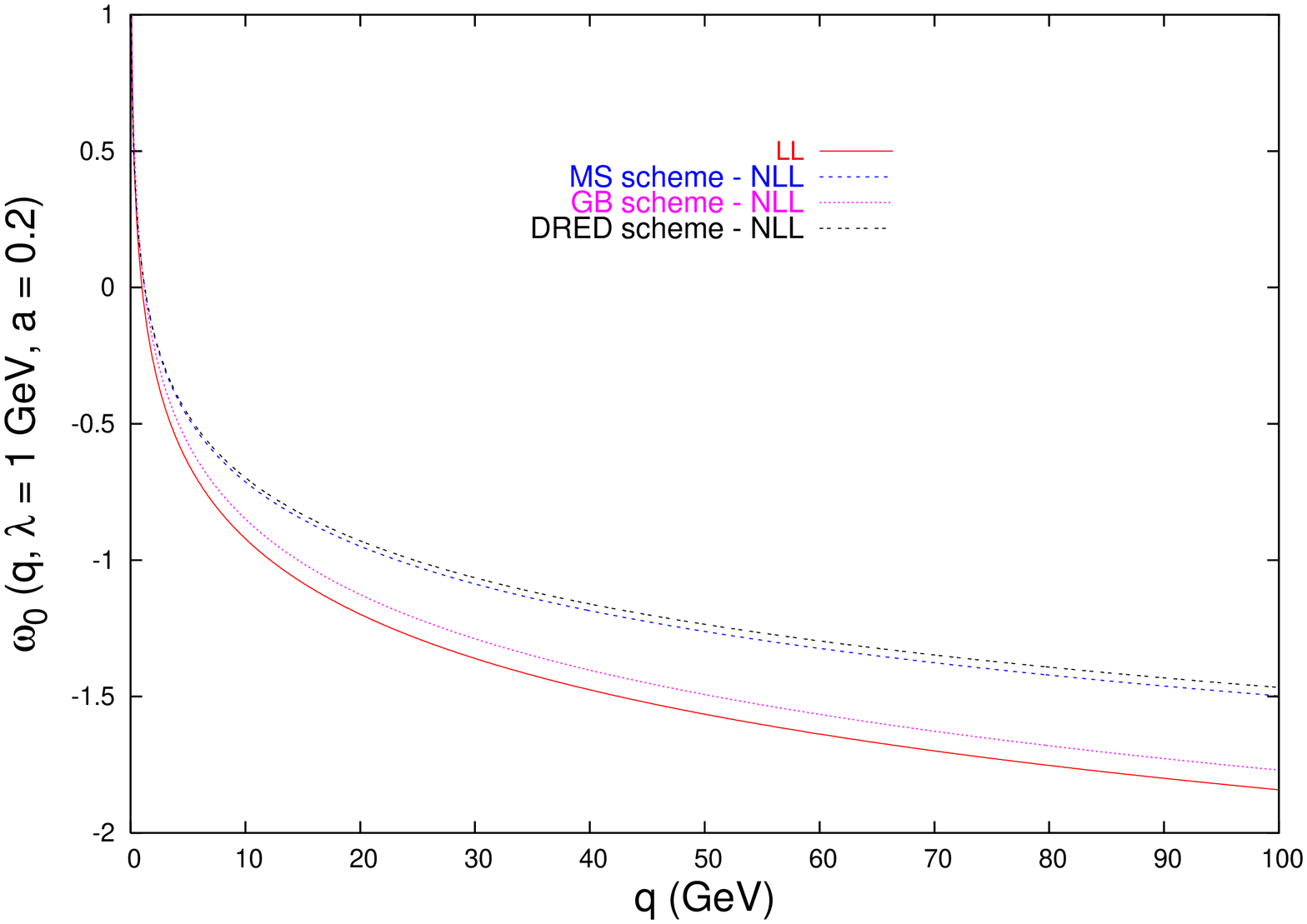,width=4.8cm,angle=-90}\psfig{file=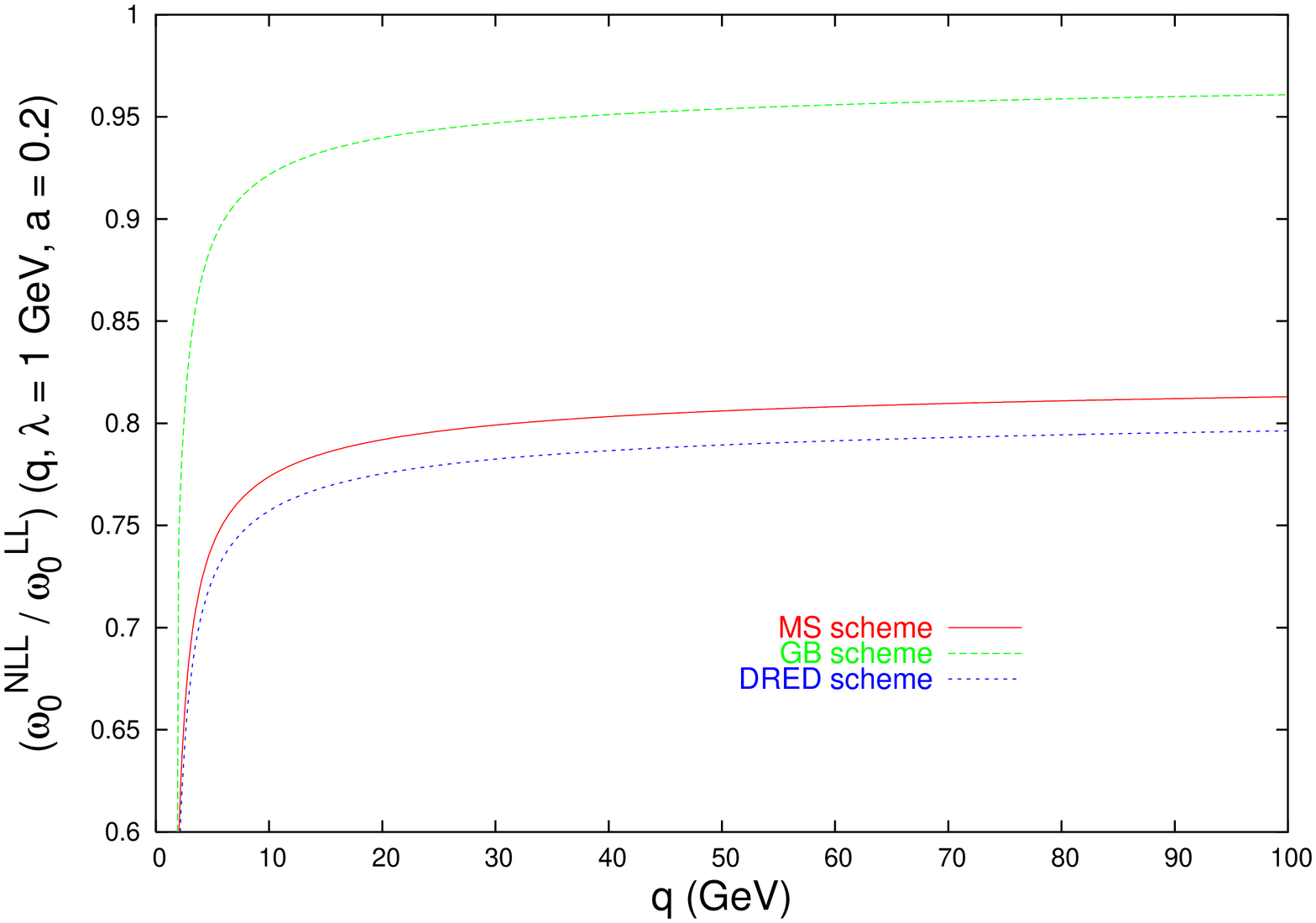,width=4.8cm,angle=-90}}
\caption{Comparison of the LL and NLL gluon Regge trajectory calculated in different schemes.}
\label{trajectory}
\end{figure}

In the following we indicate which expressions have been used in the
implementation of the solution to the BFKL equation. Firstly, in
Fig.~\ref{trajectory} it is shown the behaviour of the gluon Regge trajectory
as in Eq.~(\ref{TRAJ}), $\omega_0(\vec{q}^2,\lambda)$, as a function of
$\lambda$ and $q$. It is interesting to note that the LL trajectory always
lies below the NLL one, for all regularisation schemes. This is the opposite
effect to that found in the QCD case, see Ref.~\cite{Andersen:2003wy}.  The
correction to the trajectory is smallest in the GB scheme, with the DRED and
${\overline {\rm MS}}$ being very similar to each other.  For the $\lambda$
dependence we can see that, at a fixed value of $q = 20$ GeV, the negative
value of the trajectory decreases for lower values of $\lambda$. Due to the
logarithm, the trajectory also decreases when, for a fixed $\lambda = 1$ GeV,
$q$ increases.  For these plots the coupling was chosen to be $a = 0.2$ 
for all schemes. The plots at the right hand side of
Fig.~\ref{trajectory} show the ratio of the NLL $\omega_0(\vec{q}^2,\lambda)$
to its LL value.

\begin{figure} 
\centerline{\psfig{file=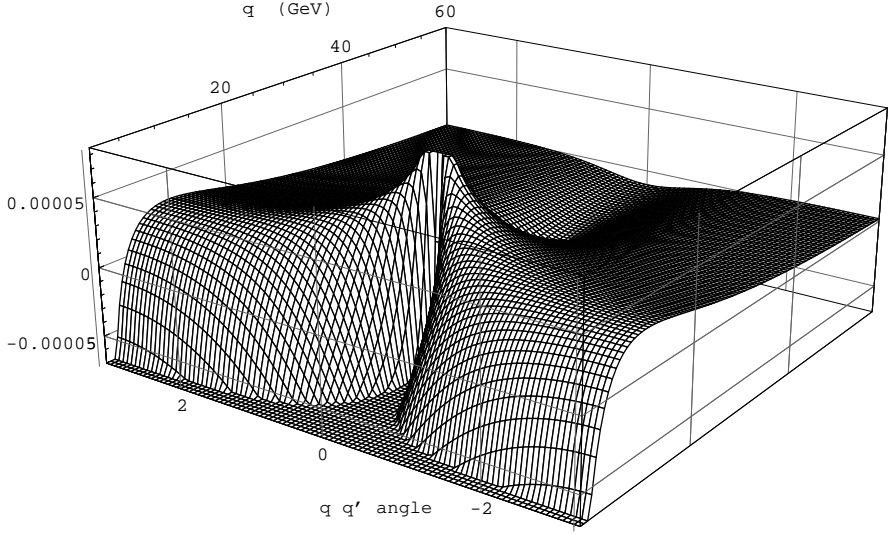,width=14cm,angle=0}}
\caption{Structure of the kernel $\tilde{K}_r \left(q,q',\theta\right)$ for 
$q' = 20$ GeV as a function of $q$ and the angle between $\vec{q}$ and 
$\vec{q}'$ for the coupling $a=0.2$.}
\label{Kernel}
\end{figure}

Secondly, the part of the real emission kernel in Eq.~(\ref{KTILDE}), can be
written as
\begin{eqnarray}
&&\hspace{-0.6cm}\widetilde{\mathcal{K}}_r \left(q,q',\theta\right) ~=~ 
\frac{a^2}{4 \pi} \left\{ - 
\frac{1}{\left(q^2+{q'}^2-2\,q\,q'\,\cos{\theta}\right)}
\ln^2{\frac{q^2}{{q'}^2}} \right.\nonumber\\
&&\hspace{-0.2cm}
+\frac{2(q^2-{q'}^2)}{\left( (q^2 + {q'}^2)^2-4\,q^2\,{q'}^2
\,\cos^2{\theta}\right)} 
\left(\frac{1}{2}\ln{\frac{q^2}{{q'}^2}}
\ln{\frac{{q}^2 {q'}^2 \left({q}^2+{q'}^2-2\,q\,q'\,\cos{\theta}\right)^2}
{({q}^2+{q'}^2)^4}} \right.\nonumber\\
&&\hspace{1cm}\left. + \left( \int_0^{- {q}^2 / {q'}^2} -
\int_0^{- {q'}^2 / {q}^2} \right) 
dt \frac{\ln(1-t)}{t}\right) \nonumber\\
&&\hspace{-0.2cm}\left. -\frac{2\,q\,{q'}\,\left|\sin{\theta}\right|}{
\left(q^2-{q'}^2\right)^2+4\,q^2\,{q'}^2\,\sin^2{\theta}} 
\left(\mathcal{F}\left(q,q',\theta\right)
+\mathcal{F} \left(q',q,\theta\right)\right)\right\},
\end{eqnarray}
with $\theta$ being the angle between the two--dimensional vectors 
$\vec{q}$ and $\vec{q}'$. For the function 
$\mathcal{F} \left(q,q',\theta\right)$ we use the expression
\begin{eqnarray}
&&\hspace{-1cm}\mathcal{F} \left(q,q',\theta\right) ~=~ 
{\rm Im} \left\{4\,{\rm Li}_2 \left(\frac{q}{q'}\,e^{-i \left|\theta\right|}\right)- \ln{\frac{q^2}{{q'}^2}}\ln{\frac{q'\,\left|\sin{\theta}\right|-i\left(q-q'\,\cos{\theta}\right)}{q'\,\left|\sin{\theta}\right|+i\left(q-q'\,\cos{\theta}\right)}}\right\}
\end{eqnarray}
with
\begin{eqnarray}
{\rm Li}_2 \left(z\right) &=& - \int^z_0 dt \frac{\ln(1-t)}{t}.
\end{eqnarray}
As in the QCD case, see Ref.\cite{Andersen:2003wy}, this kernel has
 integrable singularities at
$\vec{q}=\vec{q}', \vec{q}=0$ and $\vec{q}'=0$. This structure is revealed when
the kernel is plotted as in Fig.~\ref{Kernel} where
$\widetilde{\mathcal{K}}_r \left(q,q' = 20 ~{\rm GeV},\theta\right)$ is shown
for $a=0.2$ in the ${\overline{\rm MS}}$
scheme.  Note how this kernel is positive in a large region of phase space,
contrary to the QCD case~\cite{Andersen:2003wy}.

\section{Study of the gluon Green's function}
\label{sec:study-gluon-greens}

\begin{figure} 
\vspace{1cm}
\centerline{\psfig{file=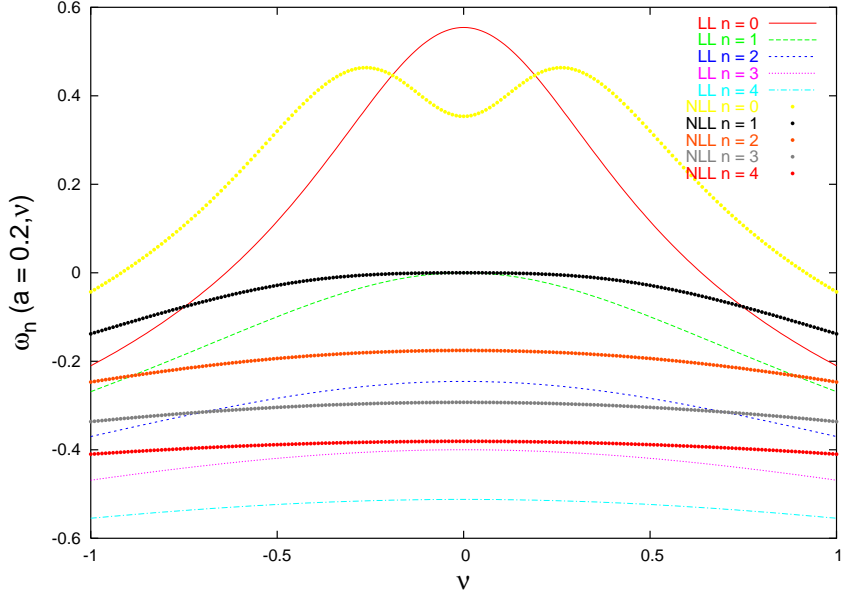,width=8cm,angle=-90}}
\caption{The eigenvalues of the BFKL kernel as a function of 
$\gamma = \frac{1}{2}+ i \nu$ at LL and NLL for different values of the 
conformal spin $n = 0,1,2,3,4$ and a coupling of $a=0.2$.}
\label{kernu}
\end{figure}

The eigenvalues of the NLL BFKL kernel as in Eq.~(\ref{eigenvaluesf}), 
$\omega_n (a, \gamma)$, along the line $\gamma =
\frac{1}{2} + i \nu$ for a value of the coupling of $a=0.2$ are plotted in
Fig.~\ref{kernu}. 
This line parameterises the contour of the $\gamma$ integration in
Eq.~(\ref{expansion}). The LL eigenvalues are compared to those obtained at
NLL for several values of the conformal spin. At high energies, for $n>0$ the
relevant region is the one close to $\nu = 0$. The zero conformal spin
evolution is governed by the two maxima and is the dominant contribution
among all conformal spins.  We will come back to this figure when the
evolution with energy of the different contributions to the gluon
Green's function is studied below.

\begin{figure} 
\vspace{1cm}
\centerline{\psfig{file=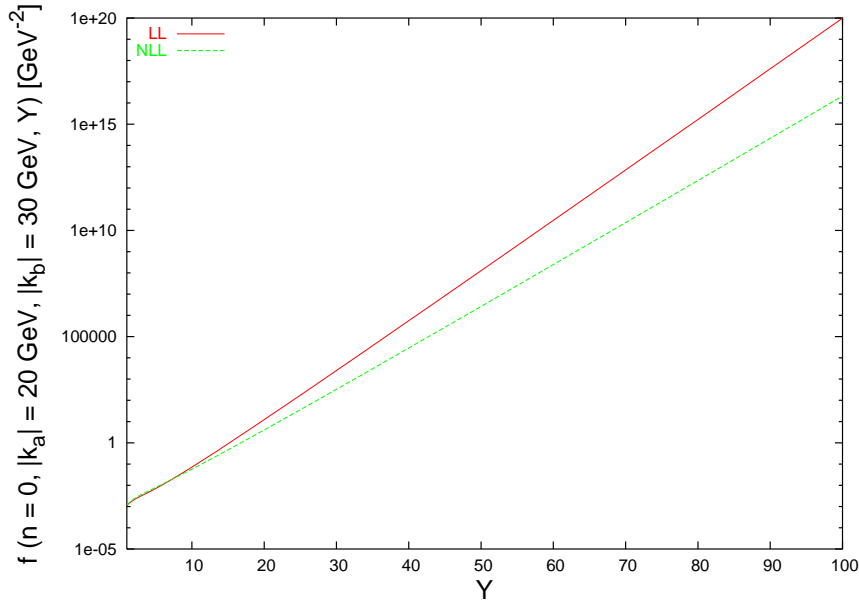,width=8cm,angle=-90}}
\caption{The gluon Green's function for zero conformal spin for fixed values 
of the external scales, a coupling of $a = 0.2$ and large values of the 
energy scale Y.}
\label{largeY}
\end{figure}

Very importantly, for a Regge--like choice of energy scale, the NLL kernel in
the N = 4 SYM theory does not develop an imaginary part along the $\gamma =
\frac{1}{2}+ i \nu$ contour. This implies that the asymptotic behaviour at
large $Y$ is well controlled and that the gluon Green's function is
  monotonically growing with energy. To illustrate this point, we
plot in Fig.~\ref{largeY} the angular averaged gluon Green's function up to
very large $Y$ both at LL and NLL accuracy.

\begin{figure} 
\vspace{1cm}
\centerline{\psfig{file=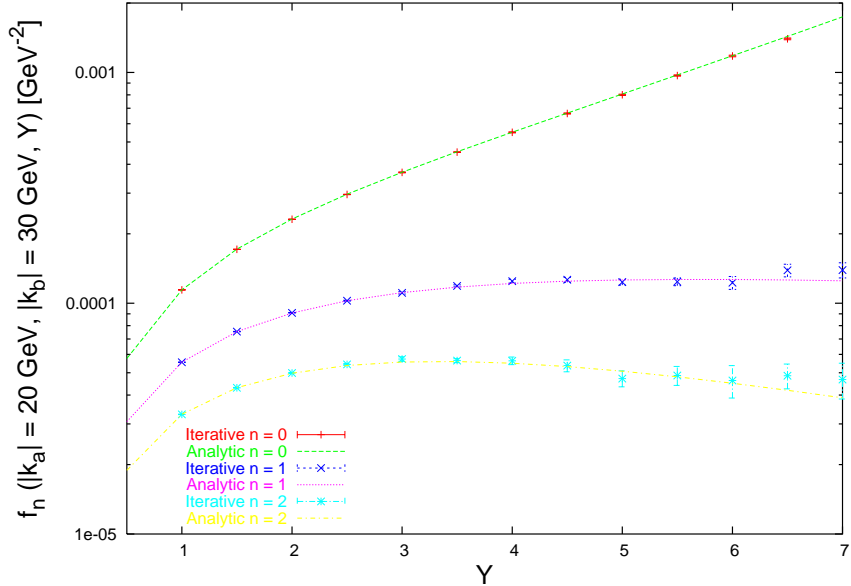,width=8cm,angle=-90}}
\caption{Contributions to the gluon Green's function from different conformal spins as a 
function of energy.}
\label{ConfSpinsY}
\end{figure}

With the intention to confirm the calculation of the conformal spins in N~=~4 
SYM of Ref.~\cite{Kotikov} the two methods of solution, that in Mellin space described in
Section~\ref{MellSol}, and the one proposed in this work directly in
transverse momentum space as explained in Section~\ref{TranSol}, will be used
to calculate the behaviour of the Green's function for different conformal
spins.  To proceed, Eq.~(\ref{expansion}) can be written as
\begin{eqnarray}
f\left(\vec{k}_a,\vec{k}_b, {\rm Y}\right)&=&
\sum_{n=-\infty}^{\infty} f_n\left(|\vec{k}_a|,|\vec{k}_b|, {\rm Y}\right) e^{i n \theta},
\label{expansion2}
\end{eqnarray}
and the coefficients in the expansion can be calculated following the 
solution in Mellin space, i.e.
\begin{eqnarray}
f_n\left(|\vec{k}_a|,|\vec{k}_b|, {\rm Y}\right) &=& 
\frac{1}{\pi |\vec{k}_a| |\vec{k}_b|} 
\int \frac{d \gamma}{2 \pi i} 
\left(\frac{\vec{k}_a^2}{\vec{k}_b^2}\right)^{\gamma-\frac{1}{2}}
e^{\omega_n (a,\gamma) {\rm Y}},
\end{eqnarray}
or they can be obtained using the solution in $\vec{k}$ space, for this 
it is needed to project on angles, i.e.
\begin{eqnarray}
f_n\left(|\vec{k}_a|,|\vec{k}_b|, {\rm Y}\right) &=& 
\int_0^{2 \pi} \frac{d\theta}{2 \pi} \, 
f\left(\vec{k}_a,\vec{k}_b, {\rm Y}\right) \cos{\left(n \theta\right)}. 
\label{nsCode}
\end{eqnarray}
Therefore the projections on conformal spins can be compared to each
other using two completely independent approaches.  The results obtained from
both solutions are shown in Fig.~\ref{ConfSpinsY}.  Both methods exactly coincide in
their predictions. This confirms the validity of the results calculated in
Ref.~\cite{Kotikov} and it is a very serious test of the method
first proposed in Ref.~\cite{Andersen:2003an,Andersen:2003wy} 
to study the gluon Green's function at NLL. As
expected from Fig.~\ref{kernu}, the dominant conformal spin is $n=0$, whose
corresponding eigenvalue is the only positive one at $\nu = 0$ in
Fig.~\ref{kernu}. The eigenvalue at $\nu=0$ for the $n=1$ conformal spin is
zero, and thus it is expected to give a constant contribution at large
energies. This behaviour is indeed observed in Fig.~\ref{ConfSpinsY}. For the rest of
conformal spins their contributions to the gluon Green's function decrease as
the available energy is larger, their eigenvalue being negative at the
vecinity of $\nu=0$.

\begin{figure} 
\vspace{1cm}
\centerline{\psfig{file=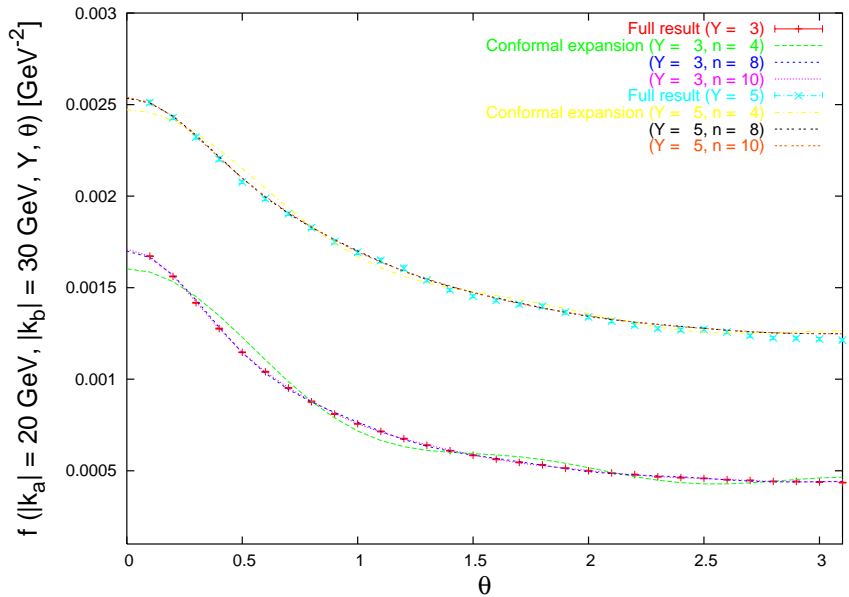,width=8cm,angle=-90}}
\caption{The dependence of the gluon Green's function on the angle between 
  the transverse momenta $\vec{k}_a$ and $\vec{k}_b$.}
\label{AllAngDep}
\end{figure}

A natural question is the convergence in $n$ of the angular expansion of
Eq.~(\ref{expansion2}). As the method of solution proposed in this paper
allows for a full determination of the angular dependence it is possible to
answer this point in a simple manner.  This issue is addressed in
Fig.~\ref{AllAngDep} where the gluon Green's function is plotted as function
of the angle between the two transverse momenta. Here it can be seen that the
conformal expansion reaches good convergence for conformal spins above $n =
8$ for the chosen values of $Y$. This plot confirms again that both
approaches produce exactly the same results. The graph is produced for two
different energies showing a stronger angular correlation for lower energies
(the curve is flatter in $\theta$ for $Y=5$), a consequence of the increasing
dominance of the zero conformal spin at larger energies.

\section{Conclusions}
The solution to the NLL BFKL equation for the N~=~4 supersymmetric Yang--Mills field 
theory has been studied in detail. In particular, a newly proposed method of solution has 
been used which allows for a detailed study of the dependence of the gluon Green's function 
on the conformal spins. The results of this work confirm that the calculations of 
Ref.~\cite{Kotikov} are correct and, simultaneously, that the proposed method of solution 
of Ref.~\cite{Andersen:2003an,Andersen:2003wy} for the NLL BFKL equation provides the 
true answer and accurate description of angular dependences in the multigluon 
ladder. The growth with energy of the gluon Green's function for the Regge--like choice 
of scale has also been demonstrated.

\noindent {\bf Acknowledgements}
We thank Tolya Kotikov and Lev Lipatov for very useful discussions, the CERN
Theory Division for hospitality, and the IPPP, University of Durham, for use
of computer resources. J.R.A. wishes to thank the II. Institut f\"ur
Theoretische Physik, University of Hamburg, and A.S.V. thanks the Cavendish Laboratory at
the University of Cambridge for hospitality.


\begin{thebibliography}{00}

\bibitem{FKL}
L.~N.~Lipatov,
%``Reggeization Of The Vector Meson And The Vacuum Singularity In Nonabelian Gauge Theories,''
Sov.\ J.\ Nucl.\ Phys.\  {\bf 23} (1976) 338
[Yad.\ Fiz.\  {\bf 23} (1976) 642],\\
%%CITATION = SJNCA,23,338;%%    
E.~A.~Kuraev, L.~N.~Lipatov and V.~S.~Fadin,
%``The Pomeranchuk Singularity In Nonabelian Gauge Theories,''
Sov.\ Phys.\ JETP {\bf 45} (1977) 199
[Zh.\ Eksp.\ Teor.\ Fiz.\  {\bf 72} (1977) 377],\\
%%CITATION = SPHJA,45,199;%%
I.~I.~Balitsky and L.~N.~Lipatov,
%``The Pomeranchuk Singularity In Quantum Chromodynamics,''
Sov.\ J.\ Nucl.\ Phys.\  {\bf 28} (1978) 822
[Yad.\ Fiz.\  {\bf 28} (1978) 1597].
%%CITATION = SJNCA,28,822;%%

\bibitem{running} L.N. Lipatov, {JETP}  ${\bf {63}}$, 904 (1986),\\
G. Camici and M. Ciafaloni,  {Phys. Lett.} B  ${\bf {395}}$, 118  (1997),\\ R.~S.~Thorne, Phys.\ Lett.\ B {\bf 474} (2000) 372, Phys.\ Rev.\ D {\bf 64} (2001) 074005,\\ J.~R.~Forshaw, D.~A.~Ross and A.~Sabio Vera, Phys.\ Lett.\ B {\bf 498} (2001) 149,\\ M.~Ciafaloni, M.~Taiuti and A.~H.~Mueller, Nucl.\ Phys.\ B {\bf 616} (2001) 349,\\ M.~Ciafaloni, D.~Colferai, G.~P.~Salam and A.~M.~Stasto, Phys.\ Lett.\ B {\bf 541} (2002) 314, Phys.\ Rev.\ D {\bf 66} (2002) 054014.

%\cite{Fadin:1998py}
\bibitem{Fadin:1998py}
V.~S.~Fadin and L.~N.~Lipatov,
%``BFKL pomeron in the next-to-leading approximation,''
Phys.\ Lett.\ B {\bf 429} (1998) 127.
%[arXiv:hep-ph/9802290].
%%CITATION = HEP-PH 9802290;%%

%\cite{Ciafaloni:1998gs}
\bibitem{Ciafaloni:1998gs}
M.~Ciafaloni and G.~Camici,
%``Energy scale(s) and next-to-leading BFKL equation,''
Phys.\ Lett.\ B {\bf 430} (1998) 349.
%[arXiv:hep-ph/9803389].
%%CITATION = HEP-PH 9803389;%%

\bibitem{NLLpapers} D.A.~Ross, Phys. Lett. {\bf B431} (1998) 161,\\
Yu.V.~Kovchegov and A.H.~Mueller, Phys. Lett. {\bf B439} (1998)
423,\\ J.~Bl\"umlein, V.~Ravindran, W.L.~van Neerven and A.~Vogt,
preprint DESY-98-036, {\tt hep-ph/9806368},\\ E.M.~Levin, preprint
TAUP 2501-98, {\tt hep-ph/9806228},\\ N.~Armesto, J.~Bartels,
M.A.~Braun, Phys. Lett. {\bf B442} (1998) 459,\\ G.P.~Salam, JHEP
{\bf 8907} (1998) 19,\\ M.~Ciafaloni and D.~Colferai, Phys. Lett.
{\bf B452} (1999) 372,\\ M.~Ciafaloni, D.~Colferai and G.P.~Salam,
Phys. Rev. {\bf D60} (1999) 114036,\\ R.S.~Thorne, Phys. Rev. {\bf
D60} (1999) 054031,\\ S.~J.~Brodsky, V.~S.~Fadin, V.~T.~Kim, L.~N.~Lipatov and G.~B.~Pivovarov, JETP Lett.\  {\bf 70} (1999) 155,\\ C.~R.~Schmidt, Phys.\ Rev.\ D {\bf 60} (1999) 074003,\\ J.~R.~Forshaw, D.~A.~Ross and A.~Sabio Vera, Phys.\ Lett.\ B {\bf 455} (1999) 273,\\ G.~Altarelli, R.~D.~Ball and S.~Forte, Nucl.\ Phys.\ B {\bf 575} (2000) 313, Nucl.\ Phys.\ B {\bf 621} (2002) 359, Nucl.\ Phys.\ B {\bf 674} (2003) 459,\\M.~Ciafaloni, D.~Colferai, G.~P.~Salam and A.~M.~Stasto, Phys.\ Lett.\ B {\bf 576} (2003) 143, Phys.\ Rev.\ D {\bf 68} (2003) 114003, Phys.\ Lett.\ B {\bf 587} (2004) 87.

%\cite{Kotikov}
\bibitem{Kotikov}
A.~V.~Kotikov and L.~N.~Lipatov, Nucl.\ Phys.\ B {\bf 582} (2000) 19, Nucl.\ Phys.\ B {\bf 661} (2003) 19 
[Erratum-ibid.\ B {\bf 685} (2004) 405].

%\cite{Kotikov:2003fb}
\bibitem{Kotikov:2003fb}
A.~V.~Kotikov, L.~N.~Lipatov and V.~N.~Velizhanin,
%``Anomalous dimensions of Wilson operators in N~=~4 SYM theory,''
Phys.\ Lett.\ B {\bf 557} (2003) 114.
%%CITATION = HEP-PH 0301021;%%

%\cite{Kotikov:2004er}
\bibitem{Kotikov:2004er}
A.~V.~Kotikov, L.~N.~Lipatov, A.~I.~Onishchenko and V.~N.~Velizhanin, hep-th/0404092.
%%CITATION = HEP-TH 0404092;%%

%\cite{Moch:2004pa}
\bibitem{Moch:2004pa}
S.~Moch, J.~A.~M.~Vermaseren and A.~Vogt,
%``The three-loop splitting functions in QCD: The non-singlet case,''
Nucl.\ Phys.\ B {\bf 688} (2004) 101.
%%CITATION = HEP-PH 0403192;%%

%\cite{Vogt:2004mw}
\bibitem{Vogt:2004mw}
A.~Vogt, S.~Moch and J.~A.~M.~Vermaseren, hep-ph/0404111.
%%CITATION = HEP-PH 0404111;%%

%\cite{Kwiecinski:1996fm}
\bibitem{Kwiecinski:1996fm}
J.~Kwiecinski, C.~A.~M.~Lewis and A.~D.~Martin,
%``Observable jets from the BFKL chain,''
Phys.\ Rev.\ D {\bf 54} (1996) 6664
%%CITATION = HEP-PH 9606375;%%

%\cite{Schmidt:1996fg}
\bibitem{Schmidt:1996fg}
C.~R.~Schmidt,
%``A Monte Carlo solution to the BFKL equation,''
Phys.\ Rev.\ Lett.\  {\bf 78} (1997) 4531
%%CITATION = HEP-PH 9612454;%%

%\cite{Orr:1997im}
\bibitem{Orr:1997im}
L.~H.~Orr and W.~J.~Stirling,
%``Dijet production at hadron hadron colliders in the BFKL approach,''
Phys.\ Rev.\ D {\bf 56} (1997) 5875
%%CITATION = HEP-PH 9706529;%%


%\cite{Andersen:2003an}
\bibitem{Andersen:2003an}
J.~R.~Andersen and A.~Sabio Vera,
%``Solving the BFKL equation in the next--to--leading approximation,''
Phys.\ Lett.\ B {\bf 567} (2003) 116.
%[arXiv:hep-ph/0305236].
%%CITATION = HEP-PH 0305236;%%

%\cite{Andersen:2003wy}
\bibitem{Andersen:2003wy}
J.~R.~Andersen and A.~Sabio Vera,
%``The gluon Green's function in the BFKL approach at next-to-leading
%logarithmic accuracy,''
Nucl.\ Phys.\ B {\bf 679} (2004) 345.
%[arXiv:hep-ph/0309331].
%%CITATION = HEP-PH 0309331;%%

\end{thebibliography}
\end{document}